\documentclass{aa}
\usepackage{graphicx}
\usepackage{txfonts}

\usepackage{natbib}

\def\Msun{M_\odot}
\def\microas{\mu{\rm as}}

\begin{document}

\title{PHASES Differential Astrometry and the Mutual Inclination of the V819 Herculis Triple Star System}

\author{Matthew W.~Muterspaugh\inst{1} \and
Benjamin F.~Lane\inst{1} \and
Maciej Konacki\inst{2, 3} \and
Bernard F.~Burke\inst{1} \and
M.~M.~Colavita\inst{4} \and
S.~R.~Kulkarni\inst{5} \and
M.~Shao\inst{4}}
\offprints{M.~Muterspaugh}

\institute{
MIT Kavli Institute for Astrophysics and Space Research, 
MIT Department of Physics, 70 Vassar Street, Cambridge, MA 02139\\
\email{matthew1@mit.edu, blane@mit.edu}
\and
Department of Geological and Planetary Sciences, California 
Institute of Technology, MS 150-21, Pasadena, CA 91125\\
\email{maciej@gps.caltech.edu}
\and
Nicolaus Copernicus Astronomical Center, 
Polish Academy of Sciences, 
Rabianska 8, 87-100 Torun, 
Poland
\and
Jet Propulsion Laboratory, California Institute of Technology, 
4800 Oak Grove Dr., Pasadena, CA 91109
\and
Division of Physics, Mathematics and Astronomy, 105-24, California 
Institute of Technology, Pasadena, CA 91125
}

\date{Received date / Accepted date}

\abstract{
V819 Herculis is a well-studied triple star system consisting of a
``wide'' pair with 5.5 year period, one component of which is a
2.2-day period eclipsing single-line spectroscopic binary.
Differential astrometry measurements from the Palomar High-precision
Astrometric Search for Exoplanet Systems (PHASES) are presented and
used to determine 
a relative inclination between the short- and long-period orbits of 
$23.6 \pm 4.9$ degrees.  
This represents only the sixth unambiguous determination of the mutual 
inclination of orbits in a hierarchical triple system.  This result is 
combined with those for the other five systems for analysis of the 
observed distribution of mutual inclinations in nearby triple systems.  
It is found that this distribution is different than that which one 
would expect from random orientations with statistical significance 
at the 94\% level; implications for studying the spatial distribution 
of angular momentum in star forming regions is discussed.
}

\authorrunning{Muterspaugh et al.}
\titlerunning{PHASES Observations of V819 Herculis}

\maketitle

\section{Introduction}

Determinations of the mutual inclinations of the two orbits in
hierarchical triple stellar systems are rare, with previously only four unambiguous
determinations available in the literature 
\citep{les93,Heintz1996,hum03}, but valuable; the dynamical relaxation
process undergone by multiples after formation is expected to leave a
statistical ``fingerprint'' in the distribution of inclinations \citep{Sterzik2002}. In
addition, well-characterized stellar multiples represent excellent
opportunities to test and challenge stellar models under stringent
constraints of common age and metallicity.

V819 Herculis (HR 6469, HD 157482; $V=5.6$, $K=4.1$) is a triple
system consisting of an evolved star (G7 III-IV; this appears brighter in V
and will be referred to as the A component) in an eccentric 5.5
year orbit together with a close ($P_n \approx 2.2$ days) pair of main
sequence F stars. The close pair (Ba and Bb) is in an edge-on orbit
and exhibits shallow eclipses. A combination of radial velocity,
speckle interferometry, eclipse timing and light-curve fitting has
made it possible to accurately determine most of the interesting
system parameters, including masses, radii and distance
\citep{Scarfe1994, vanHamme1994, Wasson1994} with accuracies of a few
percent. However, until now it has not been possible to determine the
mutual inclination of the orbits. The V819 Her system is listed as a
chromospherically active binary in the catalog by
\cite{Strassmeier1993}; it exhibits Ca H and K emission 
and has been detected in X-rays \citep{dem93} but not in radio
\citep{Drake1989}.  The system is variable with an amplitude of
approximately 80 milli-magnitudes. In addition to the eclipses 
of the close pair, the GIV 
component
exhibits quasi-periodic variability attributed
to starspots \citep{vanHamme1994}.

A technique has been developed to obtain high precision 
(10-20 $\microas$) astrometry of close stellar pairs (separation less
than one arcsecond; \citet{LaneMute2004a}) using long-baseline
infrared interferometry at the Palomar Testbed Interferometer (PTI;
\citet{col99}). This level of precision allows us to determine the orbital
parameters of the V819 Her system at the level of $\sim 3$-5\%, primarily
limited by the quality of available radial velocities.  However,
because one can measure the astrometric motion of the close pair in the system, 
it is possible to 
directly determine the relative inclination of the orbits.
These measurements were made as part of the 
Palomar High-precision Astrometric Search for Exoplanet Systems
(PHASES) program. PTI is located on Palomar Mountain near San Diego,
CA. It was developed by the Jet Propulsion Laboratory,
California Institute of Technology for NASA, as a testbed for
interferometric techniques applicable to the Keck Interferometer and
other missions such as the Space Interferometry Mission, SIM.  It
operates in the J ($1.2 \mu{\rm 
m}$), H ($1.6 \mu{\rm m}$), and K
($2.2 \mu{\rm m}$) bands, and combines starlight from two out of three
available 40-cm apertures. The apertures form a triangle with one 110
and two 87 meter baselines.

\section{Observations and Data Processing}

\subsection{PHASES Observations}

V819 Herculis was observed using PTI on 31 nights in 2003-2005 
using the observing mode described in \cite{LaneMute2004a}.  
This 
method for phase-referenced differential astrometry of 
subarcsecond binaries 
is briefly reviewed 
here.

In an optical interferometer light is collected at two or more
apertures and brought to a central location where the beams are
combined and a fringe pattern produced.  For a broadband source of central
wavelength $\lambda$ the fringe pattern appears only when the optical
paths through the arms of the interferometer are equalized to within a
coherence length ($\Lambda = \lambda^2/\Delta\lambda$). For a
two-aperture interferometer, neglecting dispersion, the intensity
measured at one of the combined beams is given by
\begin{equation}\label{double_fringe_delEqu}
I(x) = I_0 \left ( 1 + V \frac{\sin\left(\pi x/ \Lambda\right)}
{\pi x/ \Lambda} \sin \left(2\pi x/\lambda \right ) \right )
\end{equation}
\noindent where x is the differential amount of path between arms of the 
interferometer, $V$ is the fringe contrast or ``visibility'', which
can be related to the morphology of the source, and $\Delta\lambda$ is
the optical bandwidth of the interferometer assuming a flat optical
bandpass (for PTI $\Delta\lambda = 0.4 \mu$m). 

The location of the resulting interference fringes are related to the
position of the target star and the observing geometry via
\begin{equation}\label{delayEquation_delEqu}
d = \overrightarrow{B} \cdot \overrightarrow{S} + \delta_a\left(\overrightarrow{S}, t\right) + c 
\end{equation}
\noindent where d is the optical path-length one must introduce
between the two arms of the interferometer to find fringes. This
quantity is often called the ``delay.'' $\overrightarrow{B}$ is the baseline, the
vector connecting the two apertures. $\overrightarrow{S}$ is the unit vector in
the source direction, and $c$ is a constant additional scalar delay
introduced by the instrument.  The term $\delta_a\left(\overrightarrow{S},
t\right)$ is related to the differential amount of path introduced by
the atmosphere over each telescope due to variations in refractive index.
For a 100-m baseline interferometer an astrometric precision of 10
$\mu$as corresponds to knowing $d$ to 5 nm, a difficult but not
impossible proposition for all terms except that related to the
atmospheric delay.  Atmospheric turbulence, which changes over
distances of tens of centimeters, angles on order tens of arcseconds,
and on subsecond timescales, forces one to use very short exposures
(to maintain fringe contrast) and hence limits the sensitivity of the
instrument. It also severely limits the astrometric accuracy of a
simple interferometer, at least over large sky-angles.

However, in narrow-angle astrometry one is concerned with a close pair
of stars, and the observable is a differential astrometric
measurement, i.e.~one is interested in knowing the angle between the
two stars ($\overrightarrow{\Delta_s} = \overrightarrow{s_2} - \overrightarrow{s_1} $).  
The atmospheric turbulence is correlated over
small angles.  If the measurements of the two stars are simultaneous, or nearly 
so, the atmospheric term subtracts out.
Hence it is still possible to obtain high precision
``narrow-angle'' astrometry 
\citep{col94}.

To correct for time-dependent fluctuations in the atmospheric
turbulence, observations consisted of operating PTI in a
phase-referenced observing mode.  After movable mirrors in the
beam-combining lab apply delay compensation to the light collected
from two 40 cm apertures, the light from each aperture is split using
30/70 beamsplitters.  Seventy percent of the light is sent to the
phase-tracking ``primary'' interferometric beam combiner which
measures the time-dependent phase of one star's interferogram
(fringes) caused by the atmospheric turbulence, and used in a
feed-back loop to control the optical delay lines.

The other $30\%$ of the light is diverted to the ``secondary'' 
interferometric beam combiner.  In this system there is an additional 
delay line with a travel of only $\approx 500$ microns.  This is used 
to introduce delay with a sawtooth waveform with frequency on order a
Hertz.  This allows us to sample the interferograms of all stars in
the one arcsecond detector field whose projected separations are
within the scan range.  Laser metrology is used along all starlight
paths between the 30/70 split and the point of interferometric
combination to monitor internal vibrations differential to the
phase-referencing and scanning beam combiners.  For V819 Herculis, the
typical scanning rate in 2003 was one scan per second and four
intensity measurements per ten milliseconds; these values were doubled
in 2004.  The typical scan amplitude was 100 microns.  An average of
3099 scans were collected each night the star was observed over a time
span of 18 to 179 minutes.

\subsection{PHASES Data Reduction}

The data reduction algorithm used was similar to that described in 
\cite{LaneMute2004a}, but now all astrometric fitting is analyzed with differential 
right ascension and declination as parameters, rather than projected
separation. First, detector calibrations (gain, bias, and background)
are applied to the intensity measurements.  Next, a grid is constructed 
in differential right ascension and declination over which to search
(in ICRS 2000.0 coordinates).  For each point in the search grid we
calculate the expected differential delay based on the interferometer
location, baseline geometry, and time of observation for each scan.
These conversions were simplified using the routines from the Naval
Observatory Vector Astrometry Subroutines C Language Version 2.0
(NOVAS-C; see \cite{novas}).  A model of a double-fringe packet is
then calculated and compared to the observed scan to derive a $\chi^2$
value as a merit of goodness-of-fit; this is repeated for each scan,
co-adding all of the $\chi^2$ values associated with that point in the
search grid.  The model fringe template is found by observing single stars, 
incoherently averaging periodograms of their interferograms, and fitting 
a sum of Gaussians to the average periodogram.  This model effective bandpass 
is Fourier transformed into delay space to create a model interferogram.  
Sample data sets have been reanalyzed with a variety of model interferograms 
and the resulting astrometric solutions vary by less than one $\microas$; 
this is largely due to the differential nature of the measurement.  
Note that in addition to the differential delay there are several
additional parameters to the double fringe packet: fringe contrast and
relative intensities as well as mean delay. These are all adjusted to
minimize $\chi^2$ on a scan-by-scan basis.  The final $\chi^2$ surface
as a function of differential right ascension and declination is thus
derived. The best-fit astrometric position is found at the minimum
$\chi^2$ position, with uncertainties defined by the appropriate
$\chi^2$ contour---which depends on the number of degrees of freedom
in the problem and the value of the $\chi^2$-minimum.

One potential complication with fitting a fringe to the data is that
there are many local minima spaced at multiples of the operating
wavelength. If one were to fit a fringe model to each scan separately
and average (or fit an astrometric model to) the resulting delays, one
would be severely limited by this fringe ambiguity (for a 110-m
baseline interferometer operating at $2.2 \mu$m, the resulting
positional ambiguity is $\sim 4.1$ milli-arcseconds). However, by
using the $\chi^2$-surface approach, and co-adding the probabilities
associated with all possible delays for each scan, the ambiguity
disappears. This is due to two things, the first being that co-adding
simply improves the signal-to-noise ratio. Second, since the
observations usually last for an hour or even longer, the associated
baseline change due to Earth rotation also has the effect of ``smearing''
out all but the true global minimum. The final $\chi^2$-surface does
have dips separated by $\sim 4.1$ milli-arcseconds from the true 
location, but any data sets for which these show up at or above the $4\sigma$ level 
are rejected.  The final astrometry measurement and related uncertainties 
are derived by fitting only the $4\sigma$ region of the surface.

The PHASES data reduction algorithm naturally accounts for
contributions from photon and read-noise.  Unmonitored phase noise
shows up by increasing the minimum value of $\chi^2$ surface.
Comparison of this value with that expected from the number of degrees
of freedom allows us to co-add the phase noise to the fit
uncertainties.

This method has been rigorously tested on both synthetic and real
data.  Data sets are divided into equal sized subsets which are
analyzed separately.  A Kolmogorov-Smirnov test shows the formal
uncertainties from the PHASES data reduction pipeline to be consistent
with the scatter between subsets.  After an astrometric solution has
been determined, one can revisit the individual scans and determine
best-fit delay separations on a scan-by-scan basis (the fringe
ambiguity now being removed).  The differential delay residuals show
normal (Gaussian) distribution, and Allan variances of delay residuals
agree with the performance levels of the formal uncertainties and show
the residuals to be uncorrelated.  It is concluded that the PHASES data
reduction pipeline produces measurement uncertainties that are
consistent with the internal scatter of the data on intra-night timescales.

\subsection{PHASES Astrometric Results}

The differential astrometry measurements are listed in
Table \ref{phasesV819HerData}, in the ICRS 2000.0 reference frame.  In
order to evaluate the night-to-night astrometric repeatability of the
data, the PHASES data were fit to a model consisting of a Keplerian
orbit representing the Ba-Bb center of light (CL) motion and a low-order
polynomial representing motion of the A-B orbit.  The minimized value
of reduced $\chi_r^2=4$, implying either that the internal
(i.e.~derived from a single night of data) uncertainty estimates are
too low by a factor of 2, or that the simple model is not appropriate for
this system.  Replacing the polynomial model for A-B with a Keplerian
does not improve the value of $\chi_r^2$ (to be expected given the
limited fraction of the A-B orbit covered by the PHASES data set).  It is
possible starspots caused astrometric jitter on this scale.  The PHASES 
uncertainties presented in this paper have been increased by a factor
of 2 to account for this discrepancy.  The rescaled (raw) median
minor- and major-axis uncertainties are 15.2 (7.6) and 363 (181)
$\microas$.  The rescaled (raw) mean minor- and major-axis
uncertainties are 19.6 (9.8) and 568 (284) $\microas$ respectively.

\begin{table*}
\centering
Table \ref{phasesV819HerData}\\
PHASES data for V819 Herculis\\
\begin{tabular}{llllllllll}
\hline
\hline
JD-2400000.5 & $\delta$RA    & $\delta$Dec  & $\sigma_{\rm min}$ & $\sigma_{\rm maj}$ & $\phi_{\rm e}$ &  $\sigma_{\rm RA}$ & $\sigma_{\rm Dec}$ & $\frac{\sigma_{\rm RA, Dec}^2}{\sigma_{\rm RA}\sigma_{\rm Dec}}$ & N \\
             &  (mas) & (mas) & ($\microas$)                  & ($\microas$)                  & (deg)     & ($\microas$)                  & ($\microas$)                  & & \\
\hline

53109.4777819 & 49.6404 & -84.4971 & 14.7 & 566.0 & 158.77 & 527.6 & 205.4 & -0.99704 & 2011 \\
53110.4800679 & 48.0940 & -84.1339 & 23.7 & 1199.8 & 159.53 & 1124.1 & 420.3 & -0.99818 & 1334 \\
53123.4555720 & 49.1857 & -85.9322 & 36.2 & 1015.6 & 162.47 & 968.5 & 307.9 & -0.99239 & 1378 \\
53130.4397626 & 48.4773 & -86.4138 & 13.1 & 411.8 & 162.95 & 393.7 & 121.4 & -0.99359 & 2537 \\
53137.4279917 & 48.3926 & -87.1400 & 27.9 & 560.7 & 164.34 & 539.9 & 153.7 & -0.98203 & 1226 \\
53145.3928594 & 48.3081 & -87.8018 & 27.5 & 633.0 & 161.59 & 600.7 & 201.6 & -0.98962 & 1673 \\
53168.3368802 & 47.0273 & -89.7517 & 30.1 & 680.1 & 162.93 & 650.2 & 201.7 & -0.98777 & 1409 \\
53172.3496326 & 47.3437 & -90.1341 & 12.6 & 339.9 & 168.29 & 332.9 & 70.1 & -0.98310 & 2560 \\
53173.3294522 & 47.1580 & -90.3605 & 16.0 & 154.8 & 33.97 & 128.7 & 87.5 & 0.97549 & 2904 \\
53181.3314386 & 46.4330 & -90.7861 & 15.0 & 349.0 & 169.71 & 343.4 & 64.1 & -0.97112 & 2795 \\
53186.3020911 & 45.6581 & -91.1217 & 36.4 & 853.0 & 166.80 & 830.5 & 198.0 & -0.98202 & 706 \\
53187.3022539 & 46.1426 & -91.2229 & 26.0 & 882.1 & 166.94 & 859.3 & 201.0 & -0.99114 & 1578 \\
53197.2663645 & 46.1848 & -92.1526 & 9.5 & 234.1 & 164.87 & 226.0 & 61.8 & -0.98714 & 5218 \\
53198.2404599 & 46.2934 & -92.2558 & 11.4 & 109.3 & 160.36 & 103.0 & 38.3 & -0.94842 & 5404 \\
53199.2897673 & 44.0250 & -91.9450 & 49.2 & 3283.0 & 171.42 & 3246.2 & 492.3 & -0.99487 & 946 \\
53208.2505295 & 46.4281 & -92.4906 & 13.2 & 362.5 & 37.67 & 287.1 & 221.8 & 0.99719 & 6558 \\
53214.2391404 & 45.6333 & -93.3432 & 10.9 & 251.7 & 169.45 & 247.5 & 47.3 & -0.97195 & 5251 \\
53215.2293360 & 45.6361 & -93.5176 & 9.5 & 221.3 & 167.53 & 216.1 & 48.7 & -0.97963 & 5723 \\
53221.2207072 & 46.2536 & -92.9732 & 17.6 & 683.9 & 38.91 & 532.3 & 429.8 & 0.99861 & 3998 \\
53228.2083438 & 45.0879 & -94.3314 & 14.4 & 201.2 & 169.45 & 197.8 & 39.5 & -0.92815 & 3180 \\
53233.1820405 & 45.0989 & -94.8462 & 12.0 & 129.5 & 167.67 & 126.5 & 30.0 & -0.91190 & 3303 \\
53234.2006462 & 44.8264 & -94.7640 & 15.2 & 75.7 & 172.74 & 75.1 & 17.9 & -0.51361 & 3701 \\
53235.2168202 & 45.2148 & -94.9186 & 17.0 & 214.3 & 176.57 & 213.9 & 21.2 & -0.60018 & 2094 \\
53236.1665478 & 44.4594 & -94.8865 & 9.5 & 156.1 & 166.59 & 151.9 & 37.4 & -0.96553 & 6684 \\
53481.5043302 & 30.4429 & -103.3356 & 22.2 & 622.2 & 38.18 & 489.3 & 385.0 & 0.99730 & 3301 \\

\hline
\end{tabular}
\caption{
PHASES data for V819 Herculis.  All quantities are in the ICRS 2000.0 reference frame.  
The uncertainty values presented in this data have all been scaled 
by a factor of 2 over the formal (internal) uncertainties within 
each given night.  Column 6, $\phi_{\rm e}$, is the angle between the major axis of the 
uncertainty ellipse and the right ascension axis, measured from increasing differential 
right ascension through increasing differential declination (the position angle of the 
uncertainty ellipse's orientation  is $90-\phi_{\rm e}$).  
The last column is the number of scans taken during a given night.  The quadrant 
was chosen such that the larger fringe contrast is designated the primary 
(contrast is a combination of source luminosity and interferometric visibility).
}
\label{phasesV819HerData}
\end{table*}

\subsection{Potential Systematic Errors}

The fractional precision of the PHASES astrometric measurements is $\sim 10^{-4}$;
at such an ambitious level there are many possible sources of systematic error 
that could appear on inter-night timescales. In particular, the system
in question exhibits two potential astrophysical sources of measurement
noise: starspots and eclipses.

\subsubsection{Starspots}

The $\approx 40$ milli-magnitude variability of V819 Herculis A has been attributed 
to starspots.  Upper limits to shifts in the CL of a star caused 
by star spots are evaluated with a model comprised of a uniform stellar disk (radius $R$) 
except for a zero-temperature (non-emitting) circular region of radius $r$ 
tangent to the edge of the stellar disk (i.e.~centered at $x=R-r$, $y=0$).  
The CL is displaced by:
\begin{eqnarray}\label{V819Her_spot_ast}
\frac{x_c}{R} & = & \frac{\int_{-R}^{R} \int_{-\sqrt{R^2-x^2}}^{\sqrt{R^2-x^2}} x {\rm d}y {\rm d}x - 
\int_{R-2r}^{R} \int_{-\sqrt{r^2-\left(x-R+r\right)^2}}^{\sqrt{r^2-\left(x-R+r\right)^2}} x {\rm d}y {\rm d}x}
{R\pi\left(R^2-r^2\right)}\nonumber\\
& = & -\frac{r^2/R^2}{1+r/R}.
\end{eqnarray}

The non-emitting spots in this model would cause 
photometric variations proportional to the 
fractional area of the stellar disk covered:
\begin{equation}\label{V819Her_spot_phot}
\frac{F}{F_o} = 1-\frac{r^2}{R^2}
\end{equation}
\noindent where $F_o$ is the star's flux when no spots are present.  
Equations \ref{V819Her_spot_ast} and \ref{V819Her_spot_phot} provide a relationship 
between the apparent astrometric and photometric shifts caused by star spots.  
For stars of typical radius 1 milli-arcsecond, the simplified model 
gives a roughly linear relationship of 0.8 $\microas$ 
of maximum astrometric shift per milli-magnitude of photometric variability.  

The largest possible astrometric shift by a star spot is given by evaluating a 
slightly different model.  In this case, the star spot fills the 
(non-circular) area from the star's edge to a chord at distance $x_o$ from the star's 
true center.  The astrometric shift is 
\begin{eqnarray}\label{V819Her_spot_ast2}
\frac{x_c}{R} & = & \frac{\int_{-R}^{x_o} \int_{-\sqrt{R^2-x^2}}^{\sqrt{R^2-x^2}} x {\rm d}y {\rm d}x}
{R \int_{-R}^{x_o} \int_{-\sqrt{R^2-x^2}}^{\sqrt{R^2-x^2}} {\rm d}y {\rm d}x}\nonumber\\
& = & - \frac{2\left(1 - \frac{x_o^2}{R^2}\right)^{3/2}}
{3 \left( \frac{\pi}{2} + \arcsin\frac{x_o}{R} + \frac{x_o}{R}\left(1 - \frac{x_o^2}{R^2}\right)^{1/2}\right)}
\end{eqnarray}
\noindent with corresponding photometric variations of 
\begin{equation}\label{spot_phot2}
\frac{F}{F_o} = \frac{1}{\pi}\left(\frac{\pi}{2} + \arcsin\frac{x_o}{R} + \frac{x_o}{R}\left(1 - \frac{x_o^2}{R^2}\right)^{1/2}\right).
\end{equation}

In the V819 Herculis system, the effect of a single, cold starspot is
maximally $\sim 25 \microas$ assuming a stellar radius of 0.8 mas.  If
multiple starspots cause the variability, this effect is reduced. In
addition, the effect of limb-darkening is to further reduce the
astrometric error.

\begin{figure}
   \resizebox{\hsize}{!}{\includegraphics{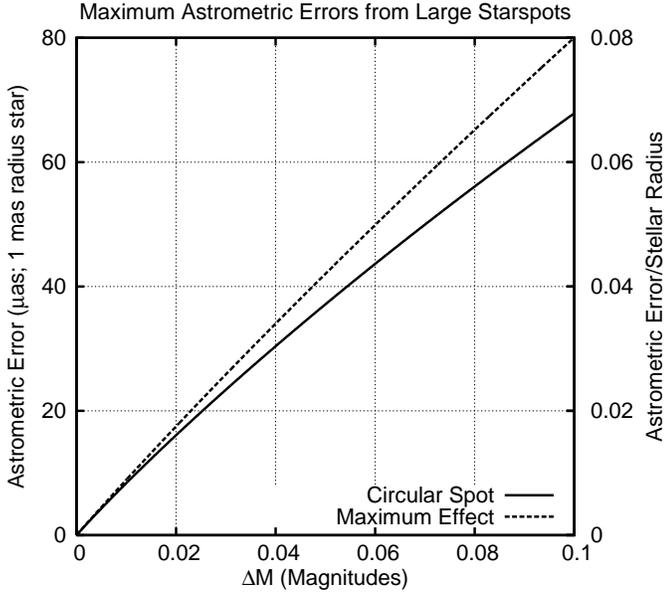}}
   \caption{
     The maximum effect of star spots on astrometric measurements 
     versus the photometric variations they cause.}
   \label{fig:v819HerStarspot_ast}
\end{figure}

\subsubsection{PHASES Observations During Ba-Bb Eclipses}

Using the published sizes and temperatures for Ba and Bb from
\cite{vanHamme1994}, it is found that the magnitude of the astrometric
shift in CL position during eclipse compared to what it
would be outside of eclipse can be greater than 100
$\microas$. Because this shift is larger than PHASES astrometric
measurement precisions, six measurements taken during eclipse are omitted 
from the data tables and the fits.

\subsection{Previous Differential Astrometry Measurements}

Previously published differential astrometry measurements of the A-B
(wide) system have been tabulated by \cite{hart04} in the Fourth
Catalog of Interferometric Measurements of Binary Stars.  In several
cases discrepancies were found between uncertainties quoted in the
original works (or uncertainty estimates omitted in the catalog); in
these cases the uncertainty estimates from the original
works are used.  All of these measurements were made using the technique of
speckle interferometry.  These measurements are included in the combined
fit to help complete coverage of the A-B orbit.

Many of the previous differential astrometry measurements were
published without any associated uncertainties.  To allow these to be
used in combined fits with other data sets, the average
uncertainties were determined as follows.  The uncertainties were initially assigned
values of 10 milli-arcseconds in separation and 1 degree in position
angle.  A (single) Keplerian model was fit to the data, and residuals
in separation and position angle were treated individually to remove
outliers and update the uncertainty estimates.
This procedure was iterated until uncertainties were
found consistent with the scatter.  A double Keplerian
model (as in eq.~\ref{V819HerCOLorbitEquation}, to allow for the Ba-Bb
subsystem) does not improve the fit; the measurements are insensitive
to this small signal.  These 22 data points have average uncertainties
of 5.92 milli-arcseconds and 0.689 degrees.

A Keplerian model was fit to the data points for which uncertainty
estimates were available to determine whether these were
systematically too large or too small, and to find outliers.  
The uncertainty estimates are found to be systematically too small; this factor
was larger in position angle than in separation.  Upon iteration, it
was found that the separation uncertainties for these 12 data points
needed to be increased by a factor of 2.43 and the position angle
uncertainties by 3.53.  Again no improvement was seen in fitting to a
double Keplerian model.

These previously published measurements are listed in Tables 
\ref{prevWithUncertV819HerData} and \ref{intWithoutUncertV819HerData}.

\begin{table*}
\centering
Table \ref{prevWithUncertV819HerData}\\
Previous astrometry with uncertainties\\
\begin{tabular}{llllll}
\hline
\hline
Year & $\rho$    & $\theta$  & $\sigma_{\rho}$ & $\sigma_{\theta}$ & Reference \\
     &  (mas)    & (deg) & (mas)           & (deg) & \\
\hline
1980.4766 & 36 & 278.1 & 4.86 & 7.06 & \cite{McA1983} \\
1980.4820 & 47 & 265.7 & 4.86 & 7.06 & \cite{McA1983} \\
1982.2924 & 107 & 154.7 & 2.43 & 1.76 & \cite{McA1997} \\
1991.3192 & 46 & 252.4 & 7.29 & 0.71 & \cite{Hrt1994} \\
1992.3104 & 48 & 123.4 & 7.29 & 0.71 & \cite{Hrt1994} \\
1992.6086 & 69 & 141.2 & 7.29 & 7.06 & \cite{Bag1999a} \\
1993.2058 & 96 & 149.9 & 7.29 & 0.71 & \cite{Hrt1994} \\
1994.7078 & 102 & 169.8 & 1.49 & 1.06 & \cite{Hrt2000a} \\
1995.3138 & 86 & 177.9 & 7.29 & 0.71 & \cite{Hrt1997} \\
1999.5681 & 115 & 164.0 & 6.90 & 3.21 & \cite{hor02} \\
1999.5681 & 109 & 160.4 & 6.90 & 3.21 & \cite{hor02} \\
2000.5563 & 101 & 175.6 & 6.90 & 3.21 & \cite{hor02} \\
\hline
\end{tabular}
\caption{
Previous differential astrometry data with published uncertainties for V819 Herculis.  
$\rho$ uncertainties have been increased by a factor of 2.43 and those for $\theta$ by a factor of 3.53.  
In many cases $\theta$ has been changed by 180 degrees from the value appearing in the original works.
}
\label{prevWithUncertV819HerData}
\end{table*}

\begin{table*}
\centering
Table \ref{intWithoutUncertV819HerData}\\
Previous astrometry without uncertainties\\
\begin{tabular}{llll||llll}
\hline
\hline
Year & $\rho$    & $\theta$  & Reference & Year & $\rho$    & $\theta$  & Reference \\
     &  (mas)    & (deg) & &      &  (mas)    & (deg) & \\
\hline
1981.4568 & 0.065 & 134.1  & \cite{McA1984a} & 1984.7009 & 0.076 & 188.3  & \cite{McA1987b} \\
1981.4678 & 0.061 & 134.2  & \cite{McA1984a} & 1985.4816 & 0.051 & 224.4  & \cite{McA1987b} \\
1981.4706 & 0.063 & 135.2  & \cite{McA1984a} & 1985.5228 & 0.047 & 228.3  & \cite{McA1987a} \\
1981.4732 & 0.062 & 135.6  & \cite{McA1984a} & 1986.4079 & 0.034 & 354.9  & \cite{Bag1989a} \\
1982.5027 & 0.109 & 155.2  & \cite{McA1987b} & 1986.6454 & 0.038 & 109.0  & \cite{Bag1989a} \\
1983.0702 & 0.112 & 163.0  & \cite{McA1987b} & 1987.2673 & 0.080 & 140.9  & \cite{McA1989} \\
1983.4202 & 0.105 & 165.8  & \cite{McA1987b} & 1988.2529 & 0.107 & 157.5  & \cite{McA1989} \\
1983.7151 & 0.107 & 171.3  & \cite{McA1987b} & 1988.6599 & 0.108 & 162.4  & \cite{McA1990} \\
1984.3732 & 0.100 & 180.2  & \cite{McA1987b} & 1989.2305 & 0.101 & 170.1  & \cite{McA1990} \\
1984.3760 & 0.088 & 181.4  & \cite{McA1987b} & 1989.7058 & 0.091 & 177.7  & \cite{Hrt1992b} \\
1984.3840 & 0.087 & 181.5  & \cite{McA1987b} & 1990.2651 & 0.074 & 190.2  & \cite{Hrt1992b} \\
\hline
\end{tabular}
\caption{
Previous interferometric differential astrometry data without published uncertainties for V819 Herculis.
The uncertainties presented in this table were determined by the scatter in the data.
For the values in this table, all uncertainties were taken to be $\sigma_{\rho} = 5.92$ mas and 
$\sigma_{\theta} = 0.689$ degrees.}
\label{intWithoutUncertV819HerData}
\end{table*}

\subsection{Radial Velocity Data}

A large number of radial velocity measurements of components A and Ba
from four observatories were reported in \cite{Scarfe1994}.
\citeauthor{Scarfe1994} indicate several measurements as outliers; 
these measurements have not been used.  There are 72
component A velocities and 50 component Ba velocities in the
McDonald/Kitt Peak data set, 70 component A and 49 component Ba
velocities in the DAO set, and 92 component A and 90 component Ba
velocities in the DDO data set.

The velocity measurements for each of three data sets presented 
were fit to double Keplerian models 
separately to determine the average velocity uncertainties
(measurements from McDonald and Kitt Peak were mixed together in the
original work, and these were analyzed together as one group).  As noted
in \cite{Scarfe1994}, the velocity precisions for component A differed
from those of Ba.  Uncertainties were derived for each data set by
fitting to a double Keplerian model and 
examining the scatter in the residuals for each
component separately.  The
uncertainty guesses were updated and the procedure iterated.  The
initial values for the component A velocities were 0.46 ${\rm
km\,s^{-1}}$ for the McDonald/Kitt Peak and DAO data sets, and 0.92
${\rm km\,s^{-1}}$ for the DDO set; for component Ba, all were set to
2 ${\rm km\,s^{-1}}$.  The average A and Ba uncertainties are 0.43
${\rm km\,s^{-1}}$ and 1.955 ${\rm km\,s^{-1}}$ for the McDonald/Kitt
Peak velocities, 0.465 ${\rm km\,s^{-1}}$ and 3.025 ${\rm km\,s^{-1}}$ for the DAO
measurements, and 1.015 ${\rm km\,s^{-1}}$ and 3.105 ${\rm
km\,s^{-1}}$ for the DDO observations.

\section{Orbital Models}

Basic models have been applied to the astrometric data.  
The simplifying assumption was made that the Ba-Bb subsystem is 
unperturbed by star A over the timescale of the observing program, 
allowing the model to be split into a wide (slow) interaction 
between star A and the center of mass (CM) of B, and the 
close (fast) interaction between stars Ba and Bb.  
The results presented in this paper result from modeling 
both the A-B and Ba-Bb motions with Keplerian orbits.

In general, one cannot simply superimpose the results of the two
orbits.  The observable in the PHASES measurements is the separation of star
A and the CL of the Ba-Bb subsystem.  Because the
CL of Ba-Bb, the CM of Ba-Bb, and the location of
star Ba are generally all unequal, a coupling amplitude must be added
to the combined model.  This coupling amplitude measures the relative
size of the semi-major axis of the Ba-Bb subsystem to that of the motion
of the CL of the Ba-Bb subsystem.  The sign of the
superposition is determined by the relative sizes of the mass and
luminosity ratios of the stars Ba and Bb.  As an example, if the
CL is located between the CM of Ba-Bb and the
location of star 
Ba, 
the motion of the CL will be in
opposite direction to the vector pointing from Ba to Bb.  For a
subsystem with mass ratio $M_{\rm{Bb}}/M_{\rm{Ba}}$ and luminosity ratio
$L_{\rm{Bb}}/L_{\rm{Ba}}$, the observed quantity is
\begin{equation}\label{V819Her3DorbitEquation}
\overrightarrow{y_{\rm{obs}}} = \overrightarrow{r_{\rm{A-B}}} - \frac{M_{\rm{Bb}}/M_{\rm{Ba}} - L_{\rm{Bb}}/L_{\rm{Ba}}}{\left(1+M_{\rm{Bb}}/M_{\rm{Ba}}\right)\left(1+L_{\rm{Bb}}/L_{\rm{Ba}}\right)}\overrightarrow{r_{\rm{Ba-Bb}}}
\end{equation}
\noindent where $\overrightarrow{r_{\rm{A-B}}}$ is the model separation pointing from star 
A to the CM of B, and $\overrightarrow{r_{Ba-Bb}}$
is the model separation pointing from star Ba to star Bb.  Including this coupling term for astrometric data is 
important when a full analysis including radial velocity data is made.

Alternatively, one can directly combine a model of the A-B system 
with a model of the motion of the CL of Ba-Bb.  
For purely astrometric data such a model is appropriate.  In this case, 
there is no sign change for the Ba-Bb CL model, 
and no extra coupling amplitude is required. This solution can
be used to provide a high-precision apparent orbit which 
can be used to compare different instruments.  
This model differs from 
the previous one in that the true semi-major axis of the Ba-Bb orbit 
is not a parameter, but is replaced by the apparent 
photocentric motion directly.  
\begin{equation}\label{V819HerCOLorbitEquation}
\overrightarrow{y_{\rm{obs}}} = \overrightarrow{r_{\rm{A-B}}} + \overrightarrow{r_{\rm{Ba-Bb, C.O.L.}}}
\end{equation}

\section{Orbital Solution and Derived Quantities}

The best-fit combined astrometry-radial velocity orbital solution
produces a set of parameters listed in Table
\ref{V819HerOrbitModels}. The reduced $\chi_r^2$ of the combined fit
to PHASES, radial velocity, and previous differential astrometry data
is 1.33.  This combined set has 521 degrees of freedom with 20
parameters.  This value for $\chi_r^2$ is slightly higher than one
would expect, but this is likely due to the manner in which the
uncertainties had to be derived.  All parameter uncertainties have
been increased by a factor of $\sqrt{1.33}$ to reflect this
difference.  It is noted that the value for the inclination of the 
Ba-Bb pair of $79.0 \pm 3.3$ degrees agrees with 
that determined by eclipse observations 
($81.00 \pm 0.36$ for the synchronous 
and $80.63 \pm 0.33$ for the asynchronous 
solutions of \cite{vanHamme1994}).
Also presented is a table of derived parameters of direct
astrophysical interest (\ref{V819DerivedValues}).  
A fit to the astrometric data alone does not constrain
several of the orbital parameters, but does constrain apparent
semi-major axes of the wide and narrow orbits to good precision:
$a_{A-B} = 73.9 \pm 0.6$ mas and $a_{Ba-Bb} = 108 \pm 8 \microas$ 
(Ba-Bb CL orbit).

\begin{figure*}
   \centerline{\includegraphics[width=7.5cm]{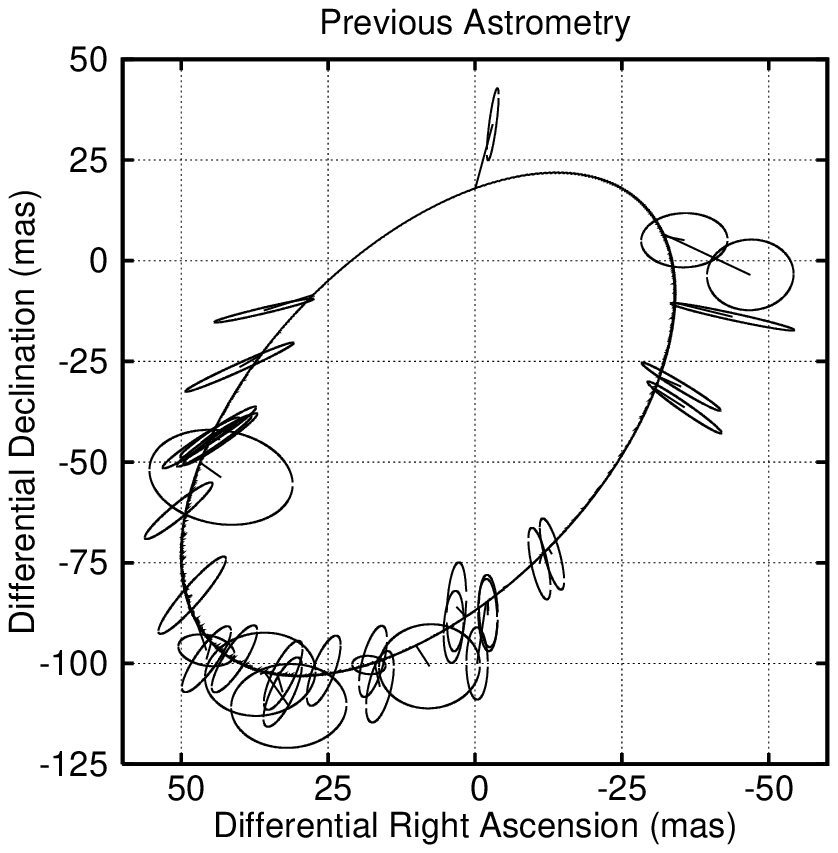}
               \includegraphics[width=7.5cm]{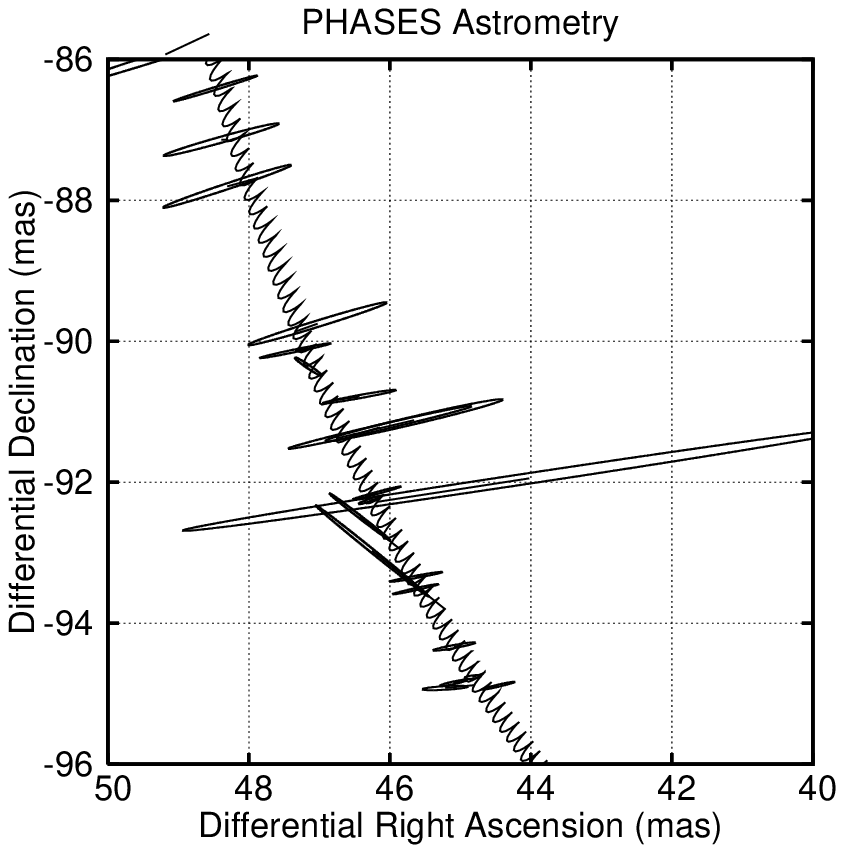}}
   \centerline{\includegraphics[width=7.5cm]{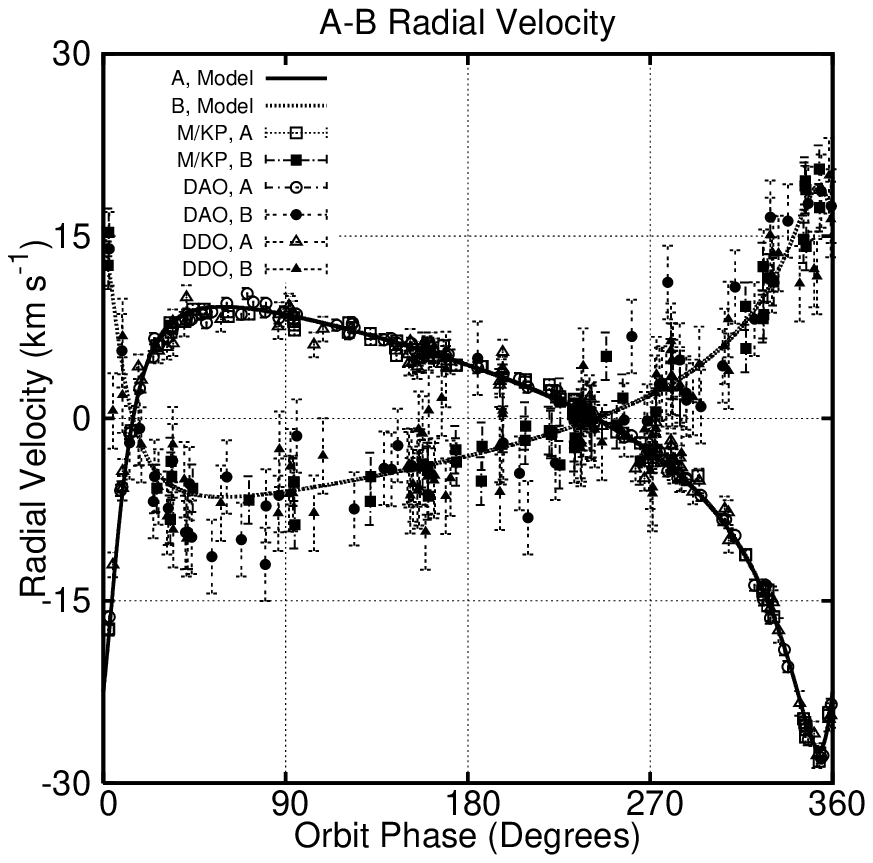}
               \includegraphics[width=7.5cm]{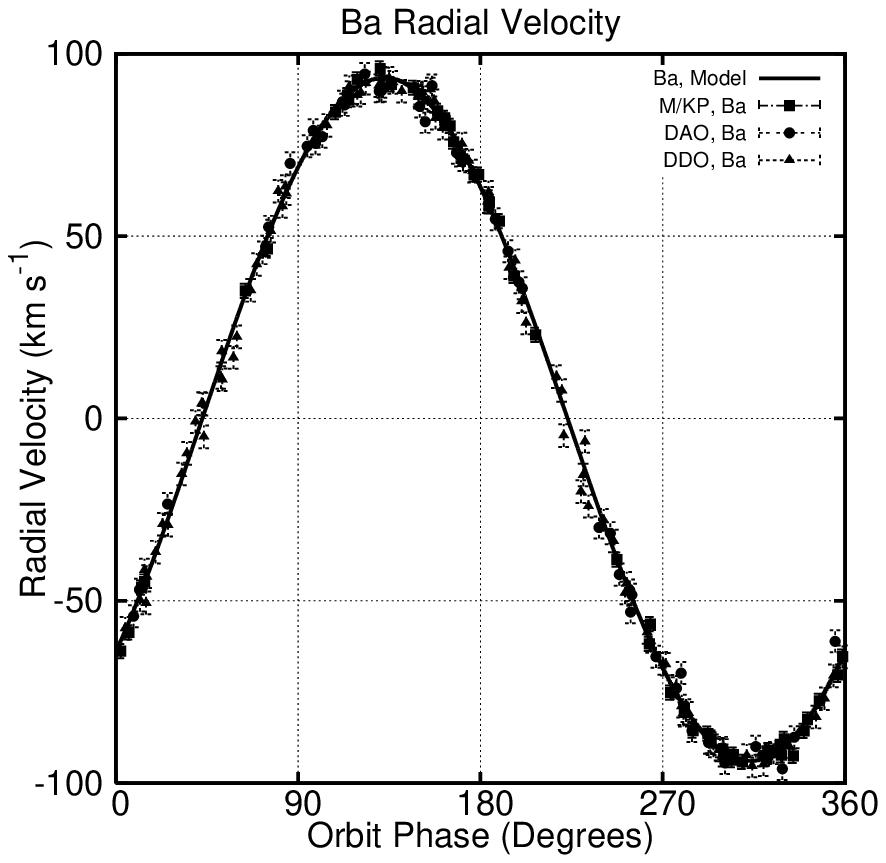}}
\caption{
The orbit of V819 Herculis.  (Top left) Previous (speckle) differential astrometry 
measurements with derived uncertainty ellipses.  (Top right) One season of PHASES
astrometry. (Bottom left) CM velocities of the wide pair. (Bottom right)
Radial velocity of the Ba component. The A-B motion has been removed
for clarity.
}
\label{v819HerOrbit}
\end{figure*}

\begin{figure*}
   \centerline{\includegraphics[width=7.5cm]{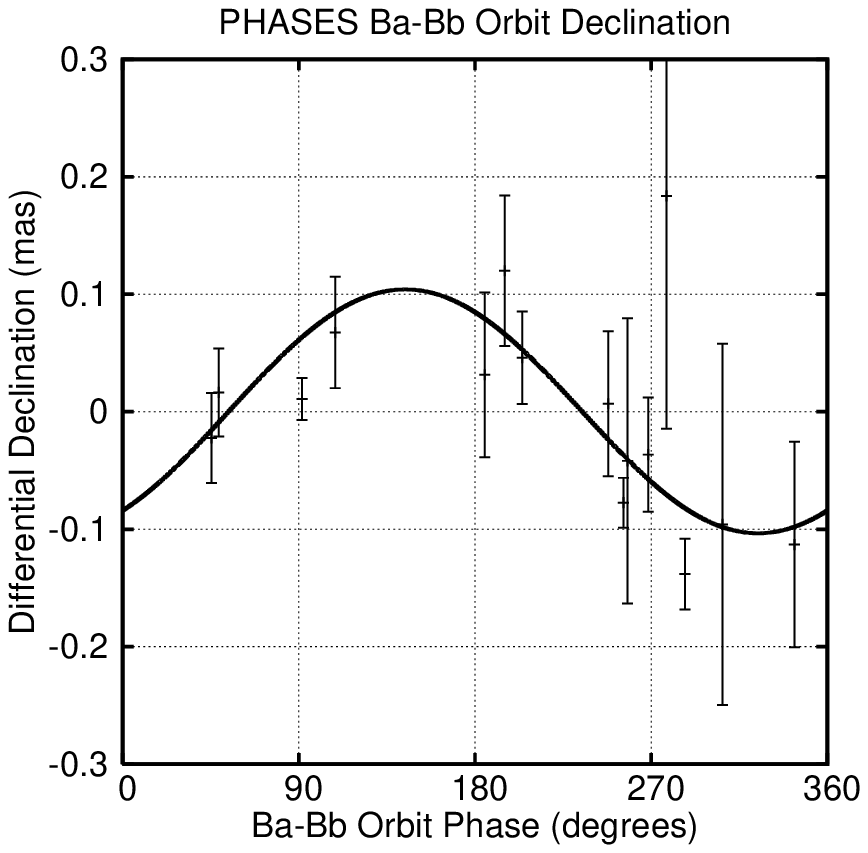}
               \includegraphics[width=7.5cm]{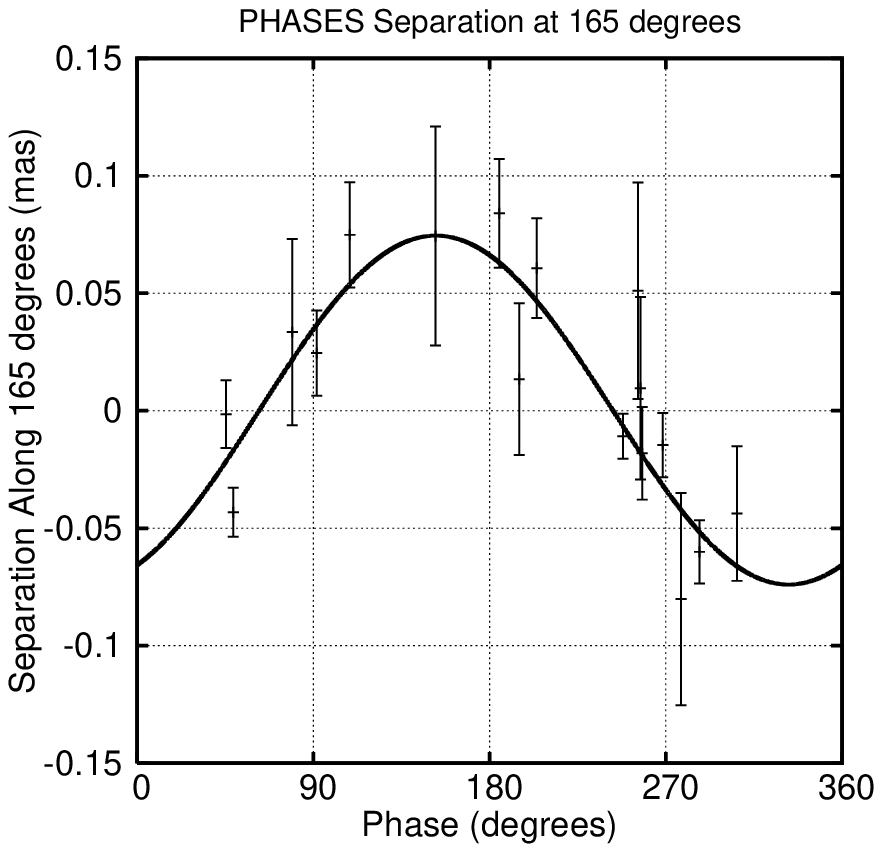}}
   \caption{CL astrometric motion of the
   V819 Herculis Ba-Bb system, as measured by PHASES observations
   along the declination axis (left) and along an axis rotated 165
   degrees East of North (equivalent to position angle 285 degrees; right); 
   this is the median position angle of
   the minor axis of the PHASES uncertainty ellipses.  Because the 
   orientation of the PHASES uncertainty ellipses varies from night to 
   night, no single axis is ideal for exhibiting the PHASES precisions, but 
   this median axis is best aligned to do so.  The wide A-B
   orbital motion has been removed for both plots.  The error bars
   plotted have been stretched by a factor of 2 over the formal
   uncertainties as discussed in the text.  The high ellipticity of
   the uncertainty ellipses causes neither the right ascension nor the
   declination uncertainties to be near the precision of the minor
   axis uncertainties, which have median uncertainty of 15.2
   $\microas$.  For clarity, measurements with projected uncertainties larger
   than 200 $\microas$ are not shown in the plots. Zero
   phase indicates periastron passage.
}
\label{v819HerPhasesBaBb}
\end{figure*}

\begin{figure*}
   \centerline{\includegraphics[width=7.5cm]{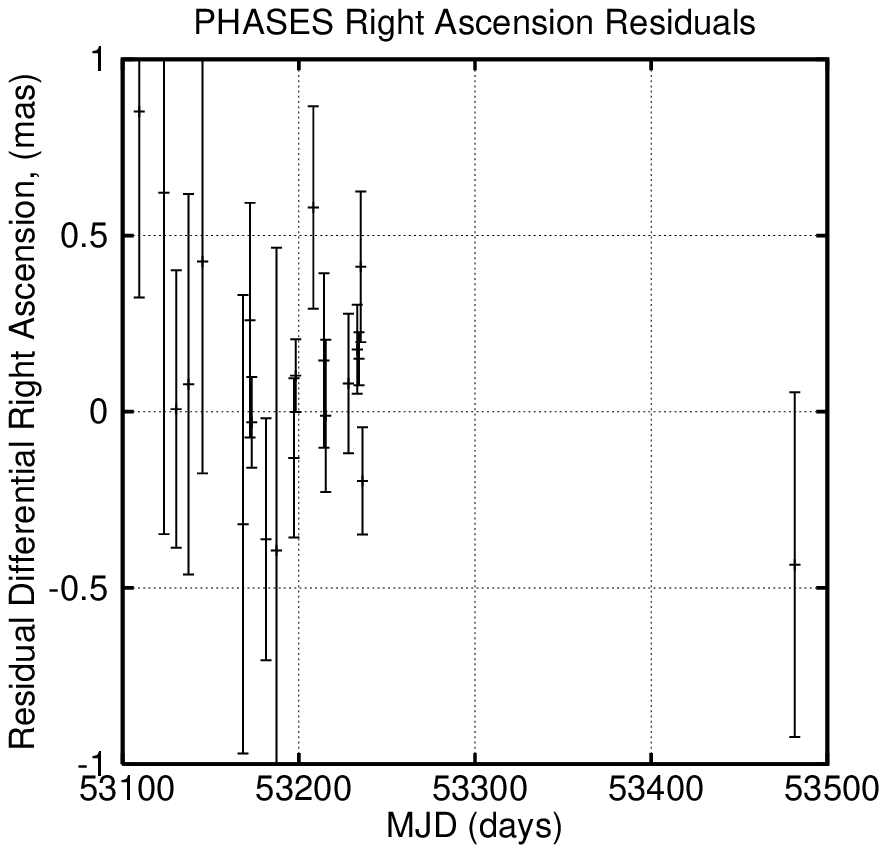}
               \includegraphics[width=7.5cm]{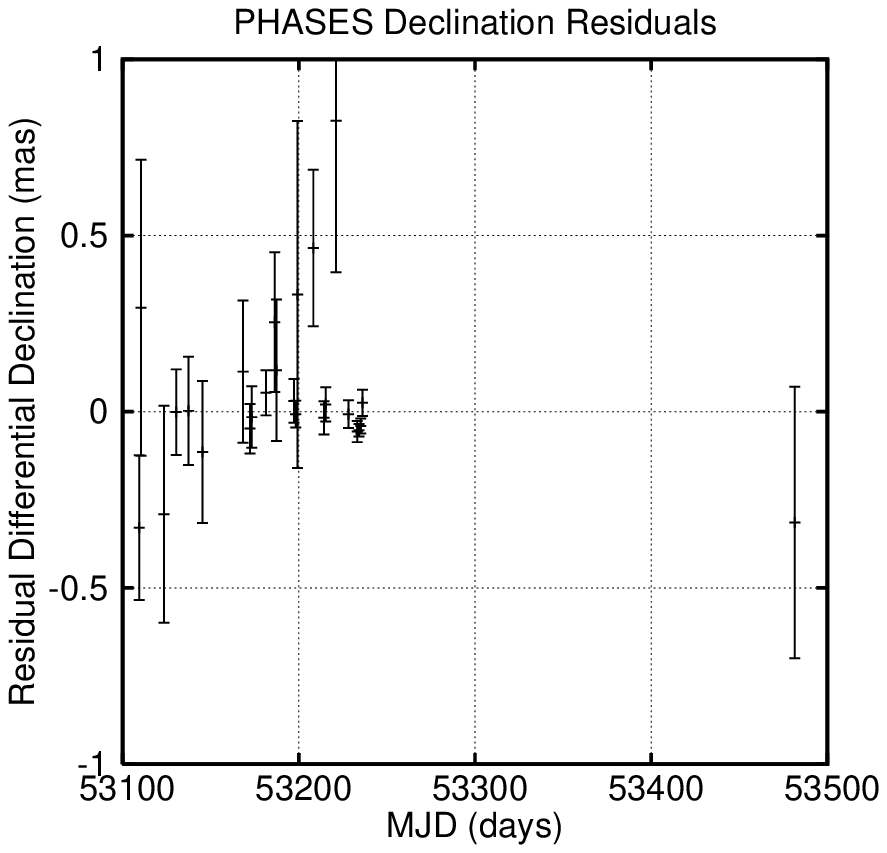}}
   \centerline{\includegraphics[width=7.5cm]{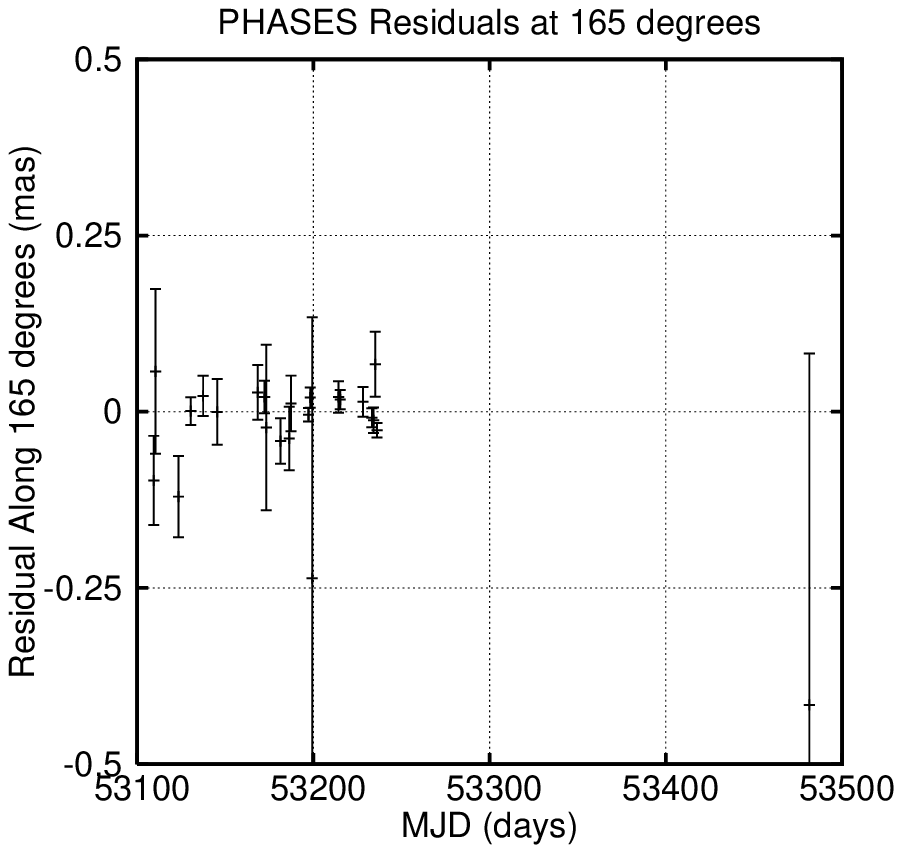}
               \includegraphics[width=7.5cm]{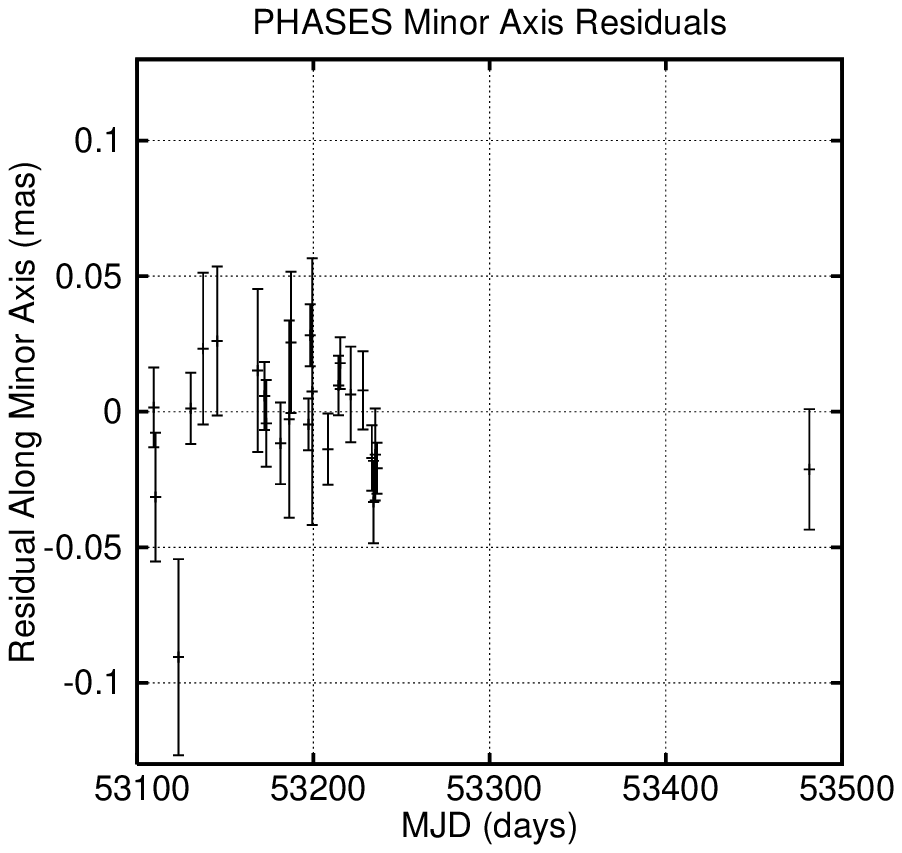}}
   \caption{
   Residuals for PHASES
   differential astrometry of V819 Herculis.  The error bars plotted
   have been stretched by a factor of 2 over the formal uncertainties
   as discussed in the text.  The high ellipticity of the uncertainty
   ellipses causes neither the right ascension nor the declination
   uncertainties to be near the precision of the minor axis
   uncertainties, which have median uncertainty of 15.2 $\microas$.
   Due to the roughly North-South alignment of the baseline used for
   most of the measurements, the more sensitive axis was typically
   declination.  The right ascension and declination plots show only
   those points for which the projected error bar is less than 1
   milli-arcsecond.  The bottom left plot shows the residuals along a
   direction that is 165 degrees from increasing differential right
   ascension through increasing differential declination 
   (equivalent to position angle 285 degrees), which
   corresponds to the median direction of the minor axis of the PHASES
   uncertainty ellipses; only measurements with uncertainties less
   than 500 $\microas$ along this axis are plotted.  Because the orientation 
   of the PHASES uncertainty ellipses varies from night to night, 
   no single axis is ideal for exhibiting the PHASES precisions, but this 
   median axis is best aligned to do so.  The bottom right
   plot shows residuals along the minor axis of each measurement's
   uncertainty ellipse.  
}
\label{v819HerPhasesResiduals}
\end{figure*}

\begin{figure*}
   \centerline{\includegraphics[width=7.5cm]{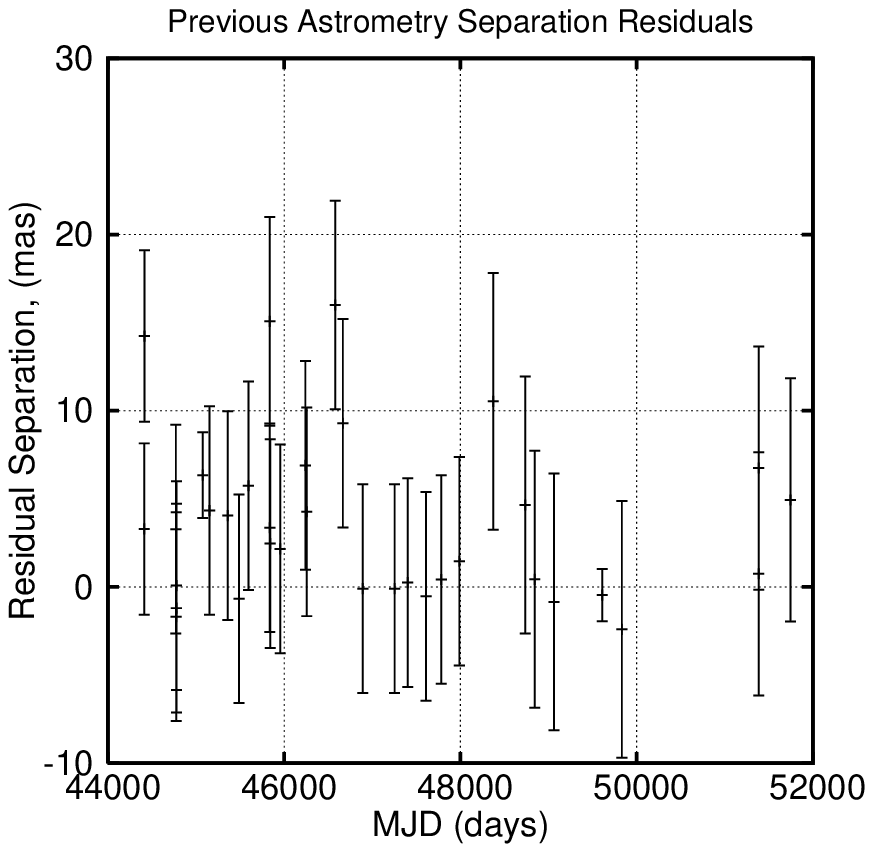}
               \includegraphics[width=7.5cm]{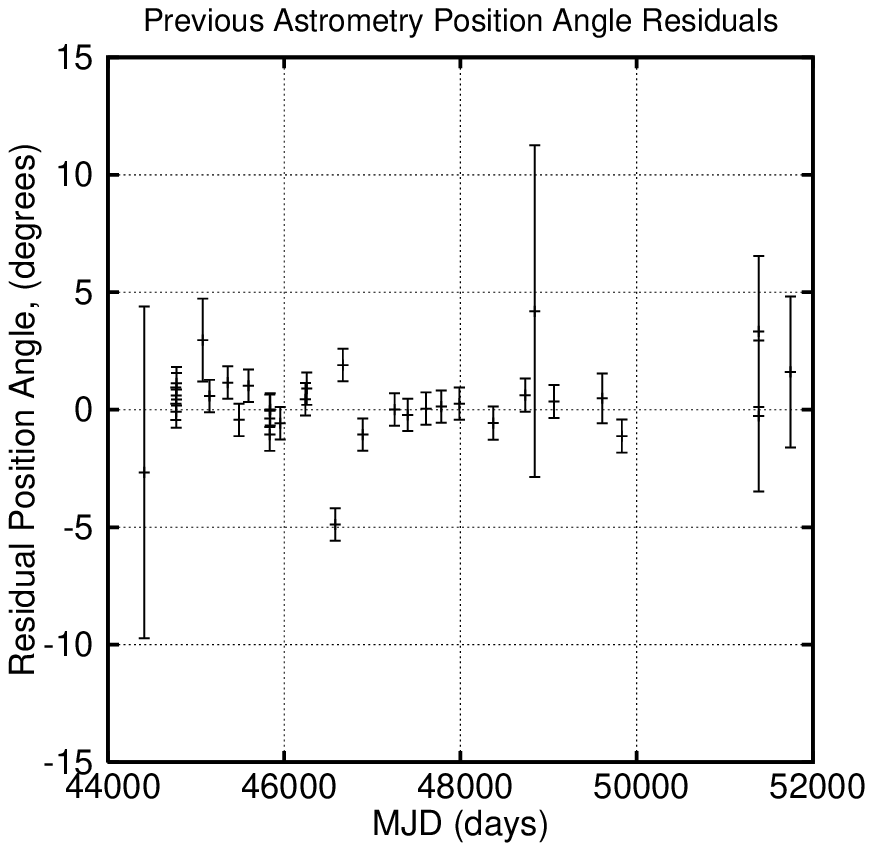}}
   \caption{
   Residuals to the combined model for previous differential astrometry of V819 Herculis.
   }
   \label{v819HerPreviousResiduals}
\end{figure*}

\begin{figure*}
   \centerline{\includegraphics[width=7.5cm]{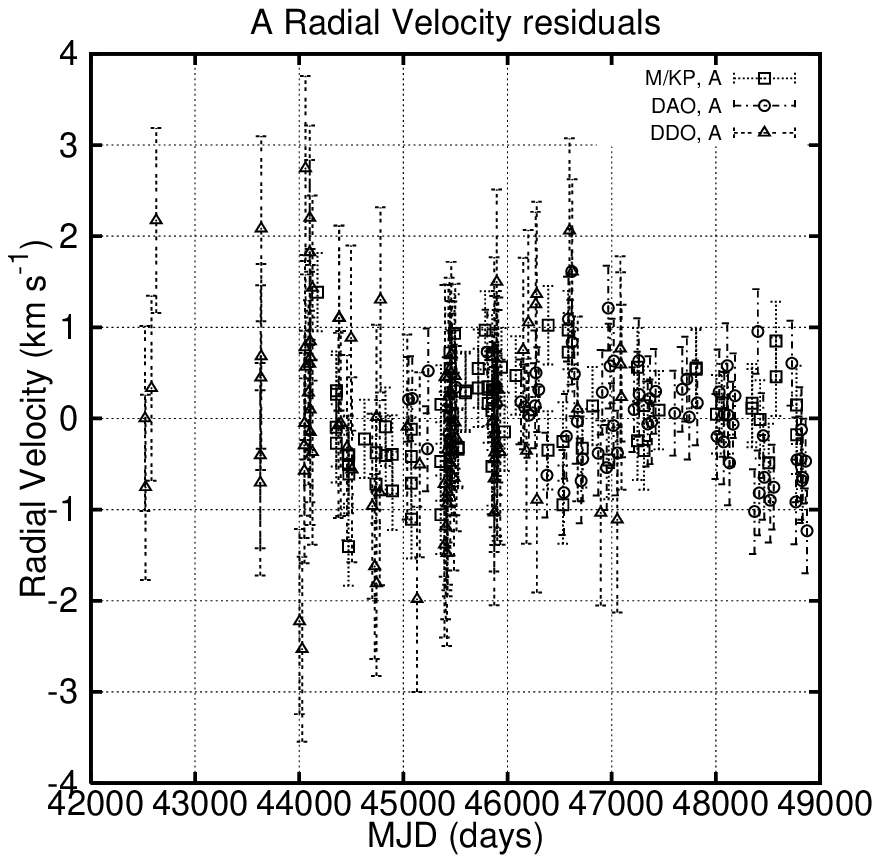}
               \includegraphics[width=7.5cm]{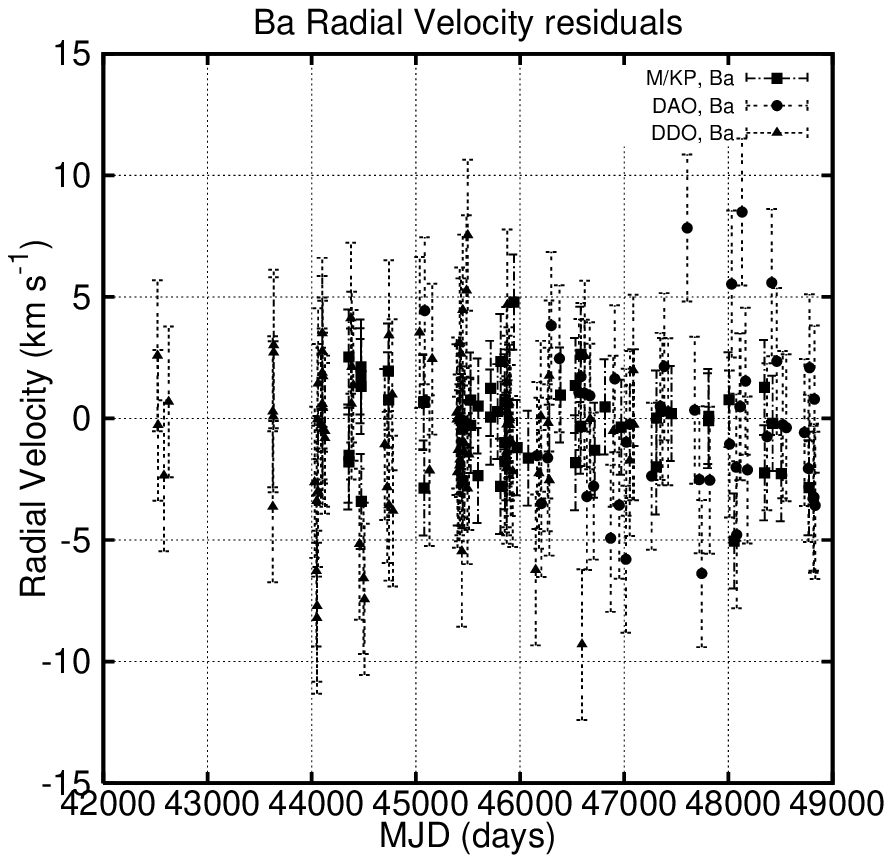}}
   \caption{
   Residuals to the combined model for radial velocimetry of V819 Herculis \citep{Scarfe1994}.
   }
   \label{v819HerRVResiduals}
\end{figure*}

The combined fit A-Ba-Bb orbit is plotted is Figure \ref{v819HerOrbit}; 
PHASES measurements of the Ba-Bb orbit CL motion is plotted in Figure 
\ref{v819HerPhasesBaBb}.  Residuals to the combined fit are shown in 
Figures \ref{v819HerPhasesResiduals}, \ref{v819HerPreviousResiduals}, 
and \ref{v819HerRVResiduals}; no evidence for additional system 
components is observed.

\begin{table}
\centering
Table \ref{V819HerOrbitModels}\\
Orbital models for V819 Herculis\\
\begin{tabular}{ll}
\hline
\hline
                                  & PHASES                          \\
                                  & $+$ Pre.~$+$ RV                 \\
\hline
$P_{A-B}$ (days)                  & 2019.79 $\pm 0.36$              \\
$T_{0, A-B}$ (MJD)                & 52628.1 $\pm 1.3$               \\
$e_{A-B}$                         & 0.6731 $\pm 0.0015$             \\
$i_{A-B}$ (degrees)               & 57.09 $\pm 0.22$                \\
$\omega_{A-B}$ (degrees)          & 42.55 $\pm 0.23$                \\
$\Omega_{A-B}$ (degrees)          & 322.40 $\pm 0.16$               \\
$P_{Ba-Bb}$ (days)                & 2.2296337 $\pm 1.9 \times 10^{-6}$ \\
$T_{0, Ba-Bb}$ (MJD)              & 52627.18 $\pm 0.30$             \\
$e_{Ba-Bb}$                       & 0.0041 $\pm 0.0033$             \\
$i_{Ba-Bb}$ (degrees)             & 79.0 $\pm 3.3$                  \\
$\omega_{Ba-Bb}$ (degrees)        & 47 $\pm 48$                     \\
$\Omega_{Ba-Bb}$ (degrees)        & 312.9 $\pm 4.8$                 \\
$V_{0, M/K}$ (${\rm km~s^{-1}}$)  & -3.388 $\pm 0.059$              \\
$V_{0, DAO}$ (${\rm km~s^{-1}}$)  & -3.373 $\pm 0.064$              \\
$V_{0, DDO}$ (${\rm km~s^{-1}}$)  & -3.35 $\pm 0.12$                \\
$M_A$ ($\Msun$)                   & 1.765 $\pm 0.095$               \\
$M_{Ba+Bb}$ ($\Msun$)             & 2.512 $\pm 0.067$               \\
$M_{Bb}/M_{Ba}$                   & 0.757 $\pm 0.020$               \\
$L_{Bb}/L_{Ba}$                   & 0.261 $\pm 0.045$               \\
$d$ (parsecs)                     & 67.96 $\pm 0.87$                \\
\hline
\end{tabular}
\caption{
Orbit models for V819 Herculis.
Pre.:  Previous differential astrometry measurements, listed in Tables 
\ref{prevWithUncertV819HerData} 
and \ref{intWithoutUncertV819HerData}.
The luminosity ratio $L_{Bb}/L_{Ba}$ is for K-band observations.
}
\label{V819HerOrbitModels}
\end{table}

\begin{table}
\centering
Table \ref{V819DerivedValues}\\
Derived physical parameters\\
\begin{tabular}{ll}
\hline
\hline
     Parameter                    & Derived Value                    \\
\hline
$a_{A-B}$ (AU)                    & $5.076   \pm 0.046$              \\ 
$a_{Ba-Bb}$ (AU)                  & $0.04541 \pm 0.00040$            \\ 
$\cos \Phi$ (degrees)             & $23.6    \pm 4.9$                \\
$M_{Ba}$ ($\Msun$)                & $1.430   \pm 0.041$              \\
$M_{Bb}$ ($\Msun$)                & $1.082   \pm 0.033$              \\
\hline
\end{tabular}
\caption{
Physical parameters for V819 Herculis derived from the combined orbital solution.
}
\label{V819DerivedValues}
\end{table}

\subsection{Mutual Inclination}

The mutual inclination $\Phi$ of two orbits is given by
\begin{equation}\label{V819HerMutualInclination}
\cos \Phi = \cos i_1 \cos i_2  + \sin i_1 \sin i_2 \cos\left(\Omega_1 - \Omega_2\right)
\end{equation}
\noindent where $i_1$ and $i_2$ are the orbital inclinations and $\Omega_1$ and $\Omega_2$ are the 
longitudes of the ascending nodes.  From the combined orbit solution a
value for this angle of $23.6 \pm 4.9 $ degrees is derived for the
V819 Her system. This low value is below the limit for
inclination-eccentricity oscillations derived by Kozai (1962; 39.2
degrees), and is consistent with the small measured value of the
eccentricity of the Ba-Bb pair.

The mutual inclination of the orbits of triple systems is 
of particular interest for understanding the conditions under 
which the system formed \citep{Sterzik2002}.  Without both radial 
velocity and visual (or astrometric) orbits for both systems in a triple, 
unambiguous determinations of the longitudes of the ascending nodes 
(and thus of the mutual inclination) are impossible.  To date there 
has only been a very small number of cases where the mutual
inclination can be unambiguously determined (Table \ref{v819herInclinations}).

\begin{table}
\centering
Table \ref{v819herInclinations}\\
Known Mutual Inclinations\\
\begin{tabular}{lll}
\hline
\hline
Star                              & Mutual Inclination   & Reference  \\
                                  &   (degrees)          & \\
\hline
V819 Her                          & $23.6 \pm 4.9 $      & This paper \\
$\kappa$ Peg                      & $43.8 \pm 3.0 $      & \citep{PHASESkapPeg_draft} \\
$\eta$ Vir                        & $30.8 \pm 1.3 $      &  \citep{hum03}  \\
$\epsilon$ Hya                    &  39.4                &  \citep{Heintz1996} \\ 
$\xi$ UMa                         &  132.1               &  \citep{Heintz1996} \\
Algol                             &  98.8 $\pm 4.9$      &  \citep{les93} \\
                                  &                      &  \citep{PanAlgol} \\
\hline
\end{tabular}
\caption{
Unambiguously known mutual inclinations of triple systems.  The value for 
Algol is determined using the measurement precisions and values 
of \cite{PanAlgol} for all but the A-B nodal position angle of 
$52 \pm 5$ degrees from \cite{les93}
}
\label{v819herInclinations}
\end{table}

\begin{figure}
   \resizebox{\hsize}{!}{\includegraphics{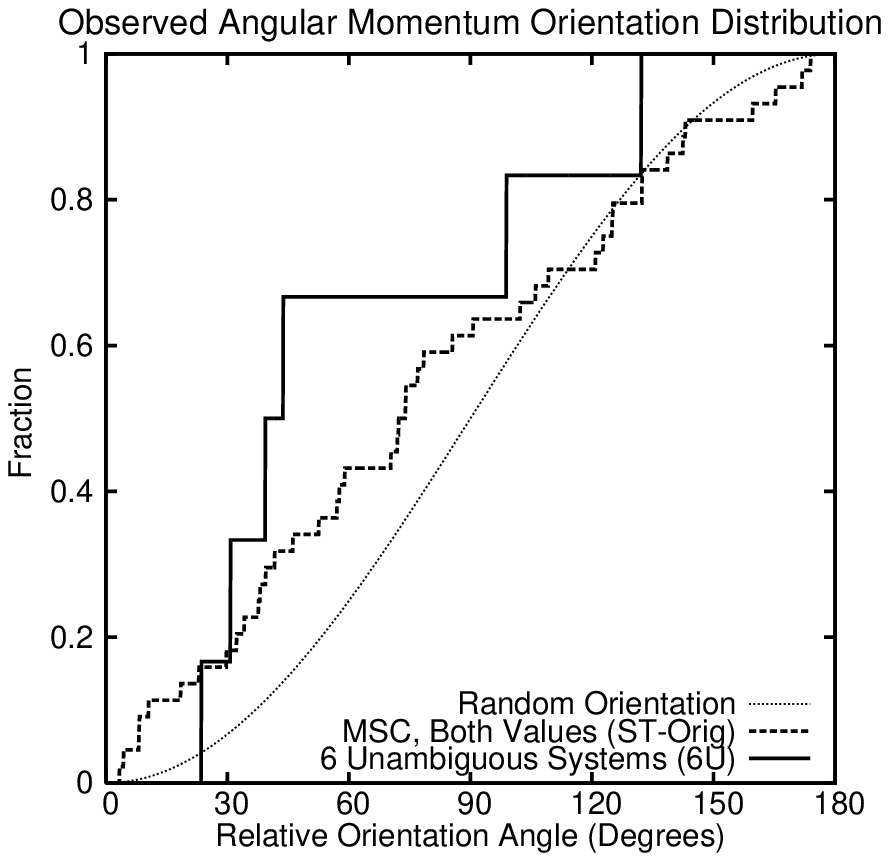}}
   \caption[Observed Angular Momentum Orientation Distribution]
    { \label{mutualInclinationCDFplot}
    Cumulative distribution of the observed distribution of angles between 
    angular momentum vectors of the six systems for which unambiguous 
    mutual inclinations have been determined.  This is compared with the 
    results from \cite{Sterzik2002}, who included 22 systems for which 
    the mutual inclinations could only be determined ambiguously---two 
    degenerate angles were both possible solutions due a 180 degree 
    ambiguity in the longitude of the ascending node of at least one 
    component of the triple system.  \citeauthor{Sterzik2002} included 
    both possible angles in their distribution.  Also shown is the 
    theoretical distribution for random orientations.
    }
\end{figure}


With the tally of systems for which unambiguous mutual inclinations have 
been determined now at six, it is reasonable to consider the distribution 
of these orientations.  The previous work on this subject is 
that of \cite{Sterzik2002}, who determined theoretical distributions resulting 
from a variety of initial conditions within star forming regions.  
At the time, the authors cited only three systems from which mutual 
inclinations were known, listed in a previous work by one of them \citep{Tok1993}.
(One of these three, $\zeta$ Cnc, is most recently listed 
by \cite{Heintz1996} as still having an ambiguous mutual inclination, 
and is not included here.)  For comparison to real 
stars, \citeauthor{Sterzik2002} instead included the 22 triple systems in 
the Multiple Star Catalog \citep{tok93msc} for which both visual orbits were 
known, but the ascending nodes had 180 degree ambiguities.  To correct for this 
lack of information, for each system they included both possible mutual inclinations 
in a combined cumulative distribution (this distribution is referred to as ST).  
This procedure assumes the ambiguity is divided evenly 
between the lower and higher possible angles; i.e.~that an equal number of the ``true'' mutual 
inclinations are the lesser of the two possible angles as are the greater.  

In Figure \ref{mutualInclinationCDFplot} the continuous distribution function 
of mutual inclinations for the six unambiguous systems 
(this distribution is referred to as 6U) is plotted with ST 
and the theoretical distribution for random orientations (referred to simply as Random).  

The two-sided Kolmogorov-Smirnov (K-S) probability for agreement 
between 6U and ST is 0.46; the one-sided KS 
probability between ST and Random is 0.07 and that 
for 6U and Random is 0.04.  Thus, the 6U set confirms the 
result of \cite{Sterzik2002} that mutual inclinations are not 
consistent with random orientations and show a slight 
preference for coplanarity.  The sets 6U and ST agree 
much better with each other than either do with random orientations, 
but the agreement probability is still low; this is likely due to the 
assumption inherent to forming the ST set by 
including both possible orientations, 
which dilutes the distribution away from coplanarity.  
A greater number of systems is 
required to better determine the distributions, and 
observational selection effects should also be considered.

\section{Conclusions}

PHASES interferometric astrometry has been used to measure the orbital
parameters of the triple star system V819 Herculis, and in particular
to resolve the apparent orbital motion of the close Ba-Bb pair. The
amplitude of the Ba-Bb CL motion is only $108 \pm 9$
$\microas$, indicating the level of astrometric precision attainable
with interferometric astrometry. By measuring both orbits one is able
to determine the mutual inclination of the two orbits, which is found to
be $23.6 \pm 4.9 $ degrees.  Such a low mutual inclination implies  
a lack of Kozai oscillations.

Further improvement in determining the system distance and component masses 
will require improved radial velocity data.  Given
that the A component is evolved and the Ba-Bb system undergoes
eclipses and hence the B components have accurately measured radii, 
this system may become a very fruitful laboratory for
high-precision testing of stellar models.

\begin{acknowledgements}

We thank the support of the PTI collaboration, whose members have
contributed designing an extremely reliable instrument for obtaining
precision astrometric measurements.  We acknowledge the extraordinary
efforts of K.~Rykoski, whose work in operating and maintaining PTI is
invaluable and goes far beyond the call of duty.  
Part of the work described in this paper was
performed at the Jet Propulsion Laboratory under contract with the
National Aeronautics and Space Administration. Interferometer data was
obtained at the Palomar Observatory using the NASA Palomar Testbed
Interferometer, supported by NASA contracts to the Jet Propulsion
Laboratory.  We thank C.~Scarfe for providing machine-readable tables
of the radial velocity measurements.  
This research has made use of the Washington Double Star Catalog 
maintained at the U.S.~Naval Observatory.
This research has made use of
the Simbad database, operated at CDS, Strasbourg, France. MWM
acknowledges the support of the Michelson Graduate Fellowship
program. BFL acknowledges support from a Pappalardo Fellowship in
Physics.  MK is supported by the NASA grant NNG04GM62G and the Polish MSIST
grant 1-P03D-021-29.  
PHASES is funded in part by the California 
Institute of Technology Astronomy Department, and by the National Aeronautics
and Space Administration under Grant No.~NNG05GJ58G issued through 
the Terrestrial Planet Finder Foundation Science Program.

\end{acknowledgements}

\object{HD 157482}
\object{HD 206901}
\object{HD 107259}
\object{HD 74874}
\object{HD 98230}
\object{HD 19356}
\object{ADS 6650}

\bibliography{main}
\bibliographystyle{plainnat}

\end{document}